\documentclass[twocolumn,seceq]{jpsj3}
\usepackage{graphicx}

\def\runauthor{
H. {\sc Ueda}, 
A. {\sc Gendiar}, 
V. {\sc Zauner}, 
T. {\sc Iharagi}, 
and T. {\sc Nishino}}

\title{Transverse Field Ising Model Under Hyperbolic Deformation}

\author
{
Hiroshi {\sc Ueda}$^{1)}$, 
Andrej {\sc Gendiar}$^{2,3)}$, 
Valentin {\sc Zauner}$^{3,4)}$, 
Takatsugu {\sc Iharagi}$^{3)}$, 
and Tomotoshi {\sc Nishino}$^{3)}$}
\inst
{
$^1$Department of Material Engineering Science, Graduate School of 
Engineering Science, Osaka University, Osaka 560-8531, Japan\\
$^2$Institute of Physics, Slovak Academy of Sciences,
D\'ubravsk\'a cesta 9, SK-845 11, Bratislava, Slovakia\\
$^3$Department of Physics, Graduate School of Science, 
Kobe University, Kobe 657-8501, Japan\\
$^4$Institut f\"ur Theoretische Physik, Technische Universit\"at Graz, A-8010 Graz, Austria
}

\abst
{
Ground state of the one-dimensional transverse field Ising model is investigated 
under the hyperbolic deformation, where the energy scale of $j$-th bond is 
proportional to the function $\cosh \bigl[ j \lambda \bigr]$ that contains a
parameter $\lambda$. Although the Hamiltonian is position dependent, the 
ground state is nearly uniform and finitely correlated. We observe the energy cross 
over between the ordered and disordered state with respect to the transverse field. 
The model shows first order phase transition, and the discontinuities in the 
magnetization and entanglement entropy at the transition point detect the Ising 
universality. 
}

\kword{DMRG, Hyperbolic, Transverse Field, Ising, Entanglement}

\begin{document}
\sloppy
\maketitle

\section{Introduction}

Spacial uniformity is a fundamental concept in physics. If a Hamiltonian of a system
is uniform, in the manner that it is represented as a spacial sum or integral of position 
independent local terms, the corresponding ground state is expected to be 
uniform, provided that there is no spontaneous symmetry breaking that causes 
spacial modulation. How about the opposite? When the ground state is uniform, 
is it expected that the Hamiltonian is also uniform? This is not true; a nonuniform 
Hamiltonian can possess a uniform ground state. A trivial example is the system 
that consists of independent spins. Consider the position dependent spin Hamiltonian
\begin{equation}
H = - \Gamma \sum_j^{~}  g_j^{~} \sigma_j^x \, ,
\end{equation}
where $\Gamma > 0$ represents the external magnetic field to the $x$-direction, 
$g_j^{~}$ a site dependent positive factor at $j$-th site, and $\sigma_j^x$ the 
Pauli operator. The corresponding 
ground state is the complete ferromagnetic state, where all the spins are polarized to 
the $x$-direction.  

The example in Eq.~(1.1) might be too trivial, since there is no inter-site couplings.
Thus let us consider the transverse field Ising (TFI) model on the 
one-dimensional (1D) lattice, as a more realistic reference system. We treat the 
position dependent TFI model defined by the Hamiltonian 
\begin{equation}
H = 
- J \sum_j^{~} f_j^{~} \sigma_j^z \sigma_{j+1}^z
- \Gamma \sum_j^{~}  g_j^{~} \sigma_j^x  \, ,
\end{equation}
where $J > 0$ represents the longitudinal nearest-neighbor coupling, and $f_j^{~}$ is 
a site dependent positive factor. When all the factors $f_j^{~}$ and $g_j^{~}$ are equal to unity, 
the ground state of this system is uniform in the thermodynamic 
limit. In this uniform case there is a quantum phase transition at $\Gamma = J$, 
where the model becomes self dual.~\cite{TFI} (See appendix.)

In this article we focus on the case where the position dependence is
given by 
$
f_j^{~} = \cosh \bigl[ j \lambda \bigr]
$
and
$
g_j^{~} = \cosh \left[ \left( j - \frac{1}{2} \right) \lambda \right] 
$, 
where $\lambda$ is a nonnegative parameter. The corresponding Hamiltonian 
\begin{eqnarray}
H^{\rm c}_{~}( \lambda ) &=&  
- J \sum_j^{~} \cosh \bigl[ j \lambda \bigr] \, \sigma_j^z \sigma_{j+1}^z \nonumber\\
&& - \Gamma \sum_j^{~} \cosh \left[ \left( j - {\textstyle \frac{1}{2}} \right) \lambda \right] \,  \sigma_j^x
\end{eqnarray}
for the case $\lambda > 0$ can be interpreted as the one parameter deformation
 --- the {\it hyperbolic deformation} --- to the uniform case when $\lambda = 0$. 
 The `bulk' part of the ground state of $H^{\rm c}_{~}( \lambda )$ is expected to be uniform 
even when $\lambda > 0$, since the path-integral representation of the imaginary
time evolution by $H^{\rm c}_{~}( \lambda )$ would be given by uniform classical
action on the $1 + 1$ hyperbolic plane.~\cite{HUeda0} Such a uniformity under hyperbolic 
deformation has been observed for the deformed $S = 1/2$ and  $S = 1$ Heisenberg spin 
chains.~\cite{HUeda1,HUeda2} We confirm this uniformity for the case of the deformed TFI model
in the next section. 

We employ the density matrix renormalization group (DMRG) 
method~\cite{White1,White2,98,Schollwoeck}, and obtain the lowest energy states 
of finite size systems under both ferromagnetic and paramagnetic boundary conditions. 
A natural interest in the ground state of $H^{\rm c}_{~}( \lambda )$ is the 
ordered-disordered transition. A classical 
analogue of the deformed TFI model is the classical Ising model on the 
two-dimensional (2D) hyperbolic lattices,~\cite{Sausset} which exhibit the 
mean-field like second-order phase transition.~\cite{Rietman,Chris1,Chris2,
dAuriac,Doyon,Shima,Hasegawa,Ueda,Roman,Roman2,Sakaniwa,
Iharagi} How about the deformed TFI 
model? We investigate the spontaneous magnetization 
in \S 3. In contrast to the classical cases, the deformed TFI model exhibits
first-order transition. The $\lambda$-dependence of the quantum entropy is 
also observed. Conclusions are summarized in the last section.

\section{Uniformity in the Ground State}

Under the hyperbolic deformation, the energy scale in the 
Hamiltonian $H^{\rm c}_{~}( \lambda )$ blows 
up exponentially with $| j |$. In order to well define the eigenvalue problem 
for $H^{\rm c}_{~}( \lambda )$, we consider 
finite-size systems. For simplicity, we treat the cases where system size $L$ is even, 
and label sites from $j = - L/2 + 1$ to $j = L/2$. Therefore the center of the system is
between $j = 0$ and $j = 1$, where the strength of the nearest neighbor interaction 
is the smallest.

Boundary condition is essential
for the determination of the ground state, since the ratio of the boundary (or
surface) energy with respect to the total energy does not vanish in the 
large $L$ limit.~\cite{Chris1,Chris2} We have to aware of this characteristic behavior of the 
hyperbolic deformation when we consider the ground-state phase transition. 
We choose either the paramagnetic boundary condition or the ferromagnetic one. 
The former is imposed by fixing the spins of both ends to the $x$-direction. 
Since the inter-site coupling in the TFI model is mediated only by longitudinal interaction 
$\sigma_j^z \sigma_{j+1}^z$, the paramagnetic boundary condition decouples the 
boundary sites at the both ends $j = - L/2 + 1$ and $j = L/2$ from the inner
part $-L/2+2 \le j \le L/2-1$. As a result, the effect of paramagnetic 
boundary condition is represented by the Hamiltonian
\begin{eqnarray}
H^{\rm c}_{{\rm P};L}( \lambda ) &=& 
- J \sum_{j = - L/2 + 2}^{j = L/2 - 2} 
\cosh \bigl[ j \lambda \bigr] \,  \sigma_j^z \sigma_{j+1}^z \\
&& - \Gamma \sum_{j = - L/2 +2}^{j = L/2 - 1} \cosh \bigl[
\left( j - {\textstyle \frac{1}{2}} \right) \lambda \bigr] \, \sigma_j^x \, , \nonumber
\end{eqnarray}
where there is no constraint for all the spins that appear in the above equation. 
On the other hand, the ferromagnetic boundary condition is imposed by fixing 
the boundary sites to the $z$-direction. In this case the Hamiltonian is expressed as
\begin{eqnarray}
\!\!\!\!\!\!\!\!\!\!\!\!
H^{\rm c}_{{\rm F};L}( \lambda ) &=& 
H^{\rm c}_{{\rm P};L}( \lambda ) \\
&+& \cosh \left[ \left( {\textstyle \frac{L}{2} -1} \right) \lambda \right] \, ( 
\sigma^z_{-L/2+2}  + 
\sigma^z_{L/2-1}
) \, , \nonumber
\end{eqnarray}
which is equivalent to put magnetic field to $z$-direction only at the position
$j = -L/2 + 2$ and $j = L/2 - 1$. Thus under both boundary conditions we
effectively treat $(L - 2)$-site system in numerical analysis. We choose the 
parameter $J$ as the unit of energy throughout this article.

We employ DMRG method~\cite{White1,White2,98,Schollwoeck} for the 
numerical determination of the ground state. A direct application of the finite-system 
DMRG algorithm encounters a  numerical instability, which is caused by the 
blow-up of the energy scale $\cosh \bigl[  j \lambda \bigr]$ with 
respect to $| j |$. In order to stabilize the computation, we treat 
\begin{eqnarray}
{\tilde H}^{\rm c}_{{\rm P};L}( \lambda ) &\equiv&
H^{\rm c}_{{\rm P};L}( \lambda ) - \left \langle
H^{\rm c}_{{\rm P};L}( \lambda ) \right \rangle \nonumber\\
{\tilde H}^{\rm c}_{{\rm F};L}( \lambda ) &\equiv&
H^{\rm c}_{{\rm F};L}( \lambda ) - \left \langle
H^{\rm c}_{{\rm F};L}( \lambda ) \right \rangle
\end{eqnarray}
instead of $H^{\rm c}_{{\rm P};L}( \lambda )$ and 
$H^{\rm c}_{{\rm F};L}( \lambda )$ directly, where $\langle ~ \rangle$ denotes
expectation value taken by the ground state. The smallest eigenvalue of both 
${\tilde H}^{\rm c}_{{\rm P};L}( \lambda )$ and 
${\tilde H}^{\rm c}_{{\rm F};L}( \lambda )$ is zero by definition. 
Since the ground state is not a priori known, the subtraction process
in Eq.~(2.3) is performed iteratively. The decay of the density matrix
eigenvalue is rapid,~\cite{Iharagi} and for all the range of deformation parameter
we examined $1/32 \le \lambda \le 1$ it is sufficient to keep $m = 16$ 
block-spin states; even for the worst case $\lambda = 1/32$ the truncation
error is of the order of $10^{-10}_{~}$ in the density matrix eigenvalues. 

\begin{figure}[h]
\centerline{\includegraphics[width=6.5cm,clip]{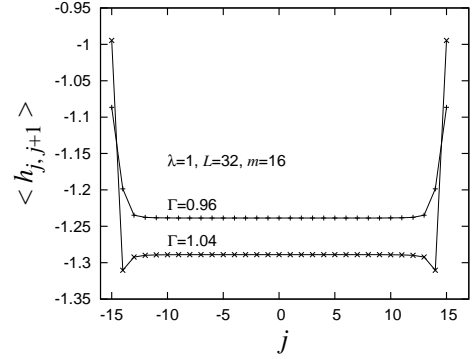}}
\caption{
Expectation value of $h_{j,j+1}^{~}$ in Eq.~(2.4) for the ground state
when $\Gamma = 0.96$ and $\Gamma = 1.04$. 
}
\end{figure}

For the analysis of of local energy, we employ a bond operator
\begin{equation}
h_{j, j+1}^{~} = - 
\left[
\frac{\Gamma}{2} \sigma_j^x + 
J \sigma_j^z \sigma_{j+1}^z +
\frac{\Gamma}{2} \sigma_{j+1}^x 
\right] \, ,
\end{equation}
which coincides with the local energy of the $j$-th bond 
of the undeformed TFI model. 
Figure 1 shows the expectation value $\langle h_{j, j+1}^{~} \rangle$ calculated 
for the ground state when $\lambda = 1$ and $L = 32$. 
We choose two typical cases, $\Gamma = 0.96$ with ferromagnetic 
boundary condition and $\Gamma = 1.04$ with paramagnetic one.
Despite of the very strong position dependence in the Hamiltonian, 
boundary correction decays rapidly, and $\langle h_{j,j+1}^{~} \rangle$ is nearly 
uniform deep inside the system. Figure 2 shows the
on-site transverse interaction $\langle \Gamma \sigma_j^x \rangle$, 
which is nearly uniform inside. We also observed similar uniformity for 
other local observables such as $\langle \sigma_j^z \rangle$ and 
$\langle \sigma_j^z \sigma_{j+1}^z \rangle$. The rapid decay of the 
boundary corrections is observed for all the values of the transverse field $\Gamma$ 
we have examined. The behavior suggests that the ground state of the 
hyperbolically deformed TFI model is always finitely 
correlated, and that the bulk part of the ground state is well approximated by the uniform
matrix product.~\cite{Ostlund1,Ostlund2}

\begin{figure}[h]
\centerline{\includegraphics[width=6.5cm,clip]{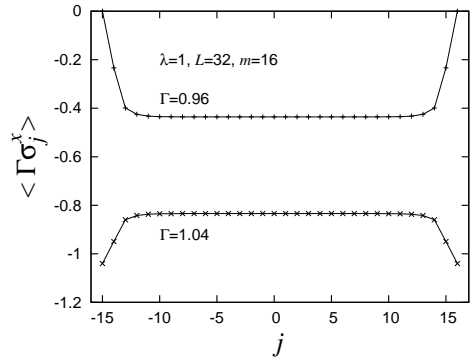}}
\caption{
On-site transverse interaction $\langle \Gamma \sigma_j^x \rangle$.
}
\end{figure}

\section{Phase Transition}

In order to capture the nature of ground-state phase transition, 
let us compare the energy of the ferromagnetic state with 
that of paramagnetic one. It is, however, difficult to directly compare
$\langle H^{\rm c}_{{\rm F};L}( \lambda ) \rangle$ with 
$\langle H^{\rm c}_{{\rm P};L}( \lambda ) \rangle$, since subtraction 
of the boundary energy is not straightforward. Therefore we use
the expectation value $\langle h_{0,1} \rangle$ at the center of the system as a 
representative value for the local energy density of the bulk. The uniformity 
observed in the previous section 
would justify this way of evaluation of the ground-state energy.
Figure 3 shows $\langle h_{0,1} \rangle$ for $L = 8$, 
$16$, $24$, and $32$ when $\lambda = 1$. Solid lines and dotted lines, 
respectively, represent data calculated with ferromagnetic and paramagnetic 
boundary conditions. System size dependence between $L = 24$ and $32$ 
is negligible, and for these cases the solid and dotted lines
crosse at $\Gamma = 1$. Although the deformed Hamiltonian is not 
invariant under the duality transformation, still the lowest 
energy state alternates at the point $\Gamma = 1$. (See Appendix.) 

\begin{figure}[h]
\centerline{\includegraphics[width=7.0cm,clip]{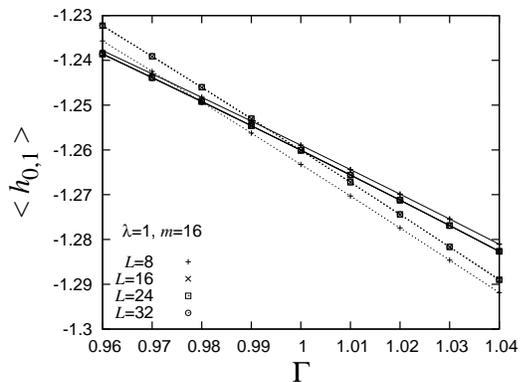}}
\caption{
Energy density $\langle h_{0,1}^{~} \rangle$ at the center of the system.
Solid lines and dotted lines, respectively, represent data calculated with 
ferromagnetic and paramagnetic boundary conditions. The $L$-dependence
is very small for $L \ge 16$.
}
\end{figure}

It should be noted that in 
a certain region of $\Gamma$ around $\Gamma = 1$ the ferromagnetic
and the paramagnetic states coexist even in the large $L$ limit, and one of 
these states is chosen by the imposed boundary condition. Thus one should be 
careful about the definition of the phase transition. Since we have considered
that $\langle h_{0,1} \rangle$ represents the bulk property of the system, 
it is natural to choose the state that gives lowest value of $\langle h_{0,1} \rangle$
as the `stable' ground state. The ground-state energy thus determined has a kink 
at the transition point $\Gamma = 1$. We checked the presence of kink in 
 $\langle h_{0,1} \rangle$ for the cases of weaker deformation down to 
 $\lambda = 1/32$, where the sufficient system size 
$L$ to obtain the convergence in  $\langle h_{0,1} \rangle$ grows roughly 
proportional to $1 / \lambda$. The fact suggests that the hyperbolic deformation 
introduces a length scale $\xi \propto 1 / \lambda$ to the system, which coincides 
the radius of curvature $R \propto - 1 / \lambda$ of the corresponding 
hyperbolic plane.~\cite{HUeda0,Sausset} Actually, the length $\xi$ already
appeared as the dumping length in Fig.~2.

\begin{figure}[h]
\centerline{\includegraphics[width=6.5cm,clip]{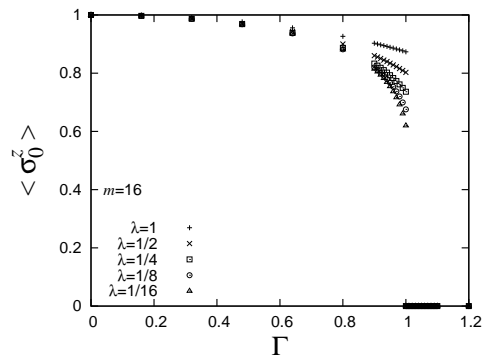}}
\caption{
Order parameter $\langle \sigma_0^z \rangle$ observed at the 
center of the system.
}
\end{figure}

Observation of the order parameter clarifies the nature of phase transition.
Figure 4 shows the magnetization $\langle \sigma_0^z \rangle$ 
at the center of the system $j = 0$, plotted for the states that gives lowest
value of $\langle h_{0,1}^{~} \rangle$; the ferromagnetic boundary
condition is imposed when $\Gamma \le 1$, and the paramagnetic condition 
when $\Gamma \ge 1$. We show the results when $\lambda = 1$, $1/2$, 
$1/4$, $1/8$, and $1/16$ calculated for sufficiently large system size $L$.
There is a finite jump in  $\langle \sigma_0^z \rangle$ at the transition 
point $\Gamma = 1$, and thus the phase transition is first order. The $\lambda$ 
dependence of this jump is plotted in Fig.~5. The eighth power of 
the jump is proportional to $\lambda$. This dependence is consistent with
the Ising universality. The length scale $\xi \propto 1 / \lambda$ introduces an 
effective deviation to the transverse field of the amount
\begin{equation}
\Delta \Gamma \propto 
\xi^{-\nu}_{~} = \frac{1}{\xi} \propto \lambda 
\end{equation}
from the criticality of the uniform TFI model, 
where $\nu = 1$ is the critical exponent for the correlation length. The observed
jump in the spontaneous magnetization is therefore proportional to  $(\Delta \Gamma )^{1/8}_{~}
\propto \lambda^{1/8}_{~}$. 

\begin{figure}[h]
\centerline{\includegraphics[width=6.5cm,clip]{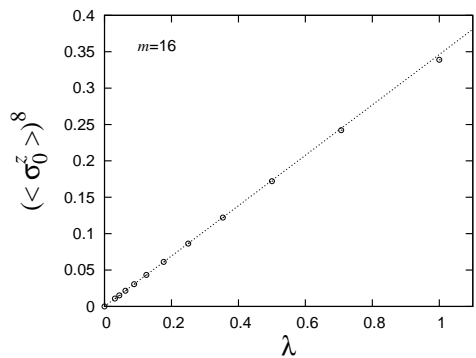}}
\caption{
Jump in the magnetization at $\Gamma = 1$ with respect to $\lambda$.
}
\end{figure}
\begin{figure}[h]
\centerline{\includegraphics[width=6.5cm,clip]{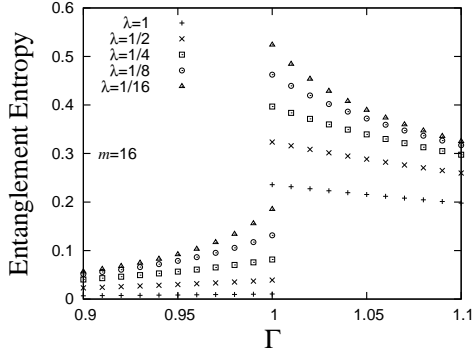}}
\caption{
Bipartite entanglement entropy measured at the center of the system.}
\end{figure}

The bipartite entanglement entropy $S$ provides supplemental information
for the phase transition. Figure 6 shows the entropy measured at the 
center of the system. Consistent with the spontaneous magnetization, 
the entropy does not show singular behavior around the  transition point. Figure 7 shows 
the entropy at $\Gamma = 1$ for the ordered state (A)  and that for the
disordered state (B), and also their sum (A)+(B) and difference (B)-(A). 
We draw a fitting line $- (1/6) \log \lambda + 0.2475$ for 
the sum (A)+(B). This suggests that the average of the
entanglement entropy [(A)+(B)]$/2$ at $\Gamma = 1$ is expressed as
\begin{equation}
\frac{1}{2} \cdot \frac{1}{6} \log \frac{1}{\lambda} + \frac{0.2475}{2} \, ,
\end{equation}
for a wide range of $\lambda$. This dependence 
coincides with the fact that the leading term of the entanglement entropy 
is expressed as $(c/6) \log W$, where $W$ is the system size and $c$ is the 
central charge.~\cite{Cardy} This is because $1 / \lambda$ is proportional to the length scale $\xi$, 
and the TFI model belongs to the class where $c = 1/2$. 

\begin{figure}[h]
\centerline{\includegraphics[width=6.5cm,clip]{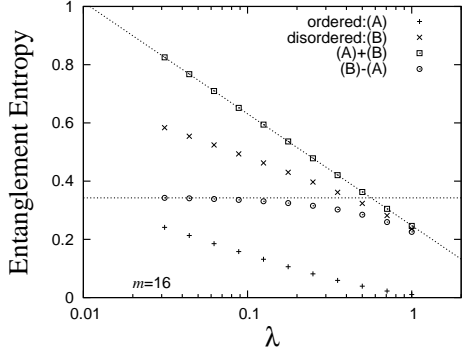}}
\caption{Entanglement entropy with respect to $\lambda$.}
\end{figure}

The fitting to the difference (B)-(A) gives the estimate $0.3426$ in the
small $\lambda$ limit. This value is about the half of $\log 2 \sim 0.6931$.
The result suggests that difference of entanglement entropy between 
ordered and disordered states is a constant, if the correlation lengths of
the both states are the same. 
We conjecture that the difference $\sim (1/2) \log 2$ captures the boundary 
condition, but the detail is not clarified yet.

\section{Conclusion and Discussion}

We have introduced the hyperbolic deformation to the 1D TFI model. It is shown
that inner part of the ground state is uniform, as was observed other systems
under hyperbolic deformation.~\cite{HUeda0,HUeda1,HUeda2} The deformation introduces a 
characteristic length $\xi$, which is proportional to $1 / \lambda$, and
the ground-state phase transition becomes 1st order. Discontinuities in both spontaneous
magnetization and bipartite entanglement entropy are consistent with the Ising 
universality class.

The observed 1st order transition is different from the mean-field like 2nd order
transition observed for the classical Ising model on hyperbolic lattices.~\cite{Rietman,
Chris1,Chris2,
dAuriac,Doyon,Shima,Hasegawa,Ueda,Roman,Roman2,Sakaniwa,Iharagi}  Thus 
spacial anisotropy may change the nature of the phase transition. To clarify the
difference, we have to construct a well defined quantum-classical 
correspondence by some means such as the Trotter decomposition.~\cite{Trotter,
Suzuki}

\section*{Acknowledgement}

This work was partly supported by Grant-in-Aid for JSPS Fellows, and 
Grant-in-Aid for Scientific Research (C) No.~22540388.
A.~G. acknowledges the support of ERDF OP R\&D, Project
hQUTE - Centre of Excellence for Quantum Technologiesh
(ITMS 26240120009), CE QUTE SAV.
% This work was partially supported by Grant-in-Aid for Scientific Research (C).

\appendix
\section{Duality Relation}

The undeformed TFI model has a symmetry mediated by the duality relation.~\cite{TFI} 
Let us introduce a new set of Pauli operators $\tau_j^x$, $\tau_j^y$, and $\tau_j^z$. 
Then the Pauli operators $\sigma_j^x$, $\sigma_j^y$, and $\sigma_j^z$ can be expressed as
\begin{eqnarray}
\sigma_j^x &=& \tau_{j-1}^z \tau_j^z \nonumber\\
\sigma_j^y &=& - \left( \prod_{\ell}^{j-2} \tau_{\ell}^x \right) \tau_{j-1}^y \tau_j^z \nonumber\\
\sigma_j^z &=& \prod_{\ell}^{j-1} \tau_{\ell}^x \, ,
\end{eqnarray}
where the transformation is nonlocal. One can verify the 
relation $\sigma_j^z \sigma_{j+1}^z = \tau_j^x$. Substituting
the above relation to $H^{\rm c}_{~}( \lambda )$ in Eq.~(1.3), we obtain the Hamiltonian
\begin{eqnarray}
H^{\rm c}_{~}( \lambda ) &=&  
- J \sum_j^{~} \cosh \bigl[ j \lambda \bigr] \,  \tau_j^z  \nonumber\\
&& - \Gamma \sum_j^{~} \cosh \left[ \left( j - {\textstyle \frac{1}{2}} \right) \lambda \right] \, \tau_{j-1}^x \tau_j^z 
\end{eqnarray}
written by the new set of Pauli operators. 
When $\lambda = 0$ the transformed Hamiltonian has the same form as the original 
Hamiltonian, where parameters $J$ and $\Gamma$ are exchanged. 
For finite $\lambda$, the transformed Hamiltonian in Eq.~(A$\cdot$2) has the form 
where lattice indices in the deformation function
is shifted by $1/2$. The observed energy crossover at $\Gamma = 1$ in Fig.~3 
suggest that this shift is not essential for the bulk part of the ground state.
If we shift the index by $1/4$ in advance and define the Hamiltonian as
\begin{eqnarray}
H^{\rm c'}_{~}( \lambda ) &=&  
- J \sum_j^{~} \cosh \left[ \left( j + {\textstyle \frac{1}{4}} \right) 
\lambda \right] \, \sigma_j^z \sigma_{j+1}^z \nonumber\\
&& - \Gamma \sum_j^{~} \left[ \cosh \left( j - {\textstyle \frac{1}{4}} \right) \lambda \right] \, \sigma_j^x
\, ,
\end{eqnarray}
then the transformed Hamiltonian is given by
\begin{eqnarray}
H^{\rm c'}_{~}( \lambda ) &=&  
- J \sum_j^{~} \cosh \left[ \left( j + {\textstyle \frac{1}{4}} \right) \lambda \right] \, \tau_j^z  \nonumber\\
&& - \Gamma \sum_j^{~} \cosh \left[ \left( j - {\textstyle \frac{1}{4}} \right) \lambda \right] \, \tau_{j-1}^x \tau_j^z \, .
\end{eqnarray}
One finds the self duality between Eq.~(A$\cdot$3) and Eq.~(A$\cdot$4) 
when $J = \Gamma$ even when $\lambda$ is positive.


\begin{thebibliography}{99}
%
\bibitem{TFI} D.C.~Mattis: {\it Theory of Magnetism II} (Springer Berlin, 1985), and references there in.
\bibitem{HUeda0}  H.~Ueda, H.~Nakano, K.~Kusakabe, and T.~Nishino: to appear in Prog. Theor. Phys; arXiv:1006.2652.
\bibitem{HUeda1} H.~Ueda and T.~Nishino: J. Phys. Soc. Jpn. {\bf 78} (2008) 014001.
\bibitem{HUeda2} H.~Ueda, H.~Nakano, K.~Kusakabe, and T.~Nishino: arXiv/0812.4513.
\bibitem{White1} S.R.~White:  Phys. Rev. Lett. {\bf 69} (1992) 2863.
\bibitem{White2} S.R.~White:  Phys. Rev. B {\bf 48} (1992) 10345.
\bibitem{98} {\it Density-Matrix Renormalization --- A new numerical method in physics ---},
eds, I.~Peschel, X.~Wang, M.~Kaulke and K.~Hallberg, (Springer Berlin, 1999), and
references there in.
\bibitem{Schollwoeck} U.~Schollw\"{o}ck: Rev. Mod. Phys. {\bf 77}   (2005) 259.
\bibitem{Sausset} F.~Sausset  and G.~Tarjus: J. Phys. A: Math. Gen. {\bf 40}   (2007) 12873.
\bibitem{Rietman} R.~Rietman, B.~Nienhuis  and J.~Oitmaa: J. Phys. A: Math. Gen. {\bf 25} (1992) 6577.
\bibitem{Chris1} N.~Anders  and C.~Chris Wu: Combinatorics, Probability and Computing {\bf 14}   (2005) 523.
\bibitem{Chris2} C.~Chris Wu: J. Stat. Phys. {\bf 100} (2000) 893.
\bibitem{dAuriac}  J.C.~Angl\'es d'Auriac, R.~M\'elin, P.~Chandra  and B.~Dou\c{c}ot: 
J. Phys. A: Math. Gen. {\bf B34} (2001) 675.
\bibitem{Doyon} B.~Doyon  and P.~Fonseca: J. Stat. Mech. (2004) P07002.
\bibitem{Shima} H.~Shima  and Y.~Sakaniwa: J. Phys. A: Math. Gen. {\bf 39} (2006) 4921.
\bibitem{Hasegawa}  I.~Hasegawa, Y.~Sakaniwa  and H.~Shima: 
Surf. Sci. {\bf 601}   (2007) 5232.
\bibitem{Ueda} K.~Ueda,  R.~Krcmar,  A.~Gendiar and  T.~Nishino: J. Phys. Soc. Jpn. {\bf 76}   (2007) 084004.
\bibitem{Roman} R.~Krcmar, A.~Gendiar, K.~Ueda and T.~Nishino: J. Phys. A Math. Theor. 
{\bf 41} (2008) 215001.
\bibitem{Roman2} R.~Krcmar, T.~Iharagi, A.~Gendiar and T.~Nishino: Phys. Rev. E {\bf 78} 
(2008) 061119.
\bibitem{Sakaniwa} Y.~Sakaniwa and H.~Shima: Phys. Rev. E {\bf 80} (2009) 021103.
\bibitem{Iharagi} T.~Iharagi, A.~Gendiar, H.~Ueda, and T.~Nishino, to appear in J. Phys. Soc.
Jpn; arXiv:1005.3378.
\bibitem{Ostlund1} S.~\"Ostlund and S.~Rommer: Phys. Rev. Lett. {\bf 75} (1995) 3537.
\bibitem{Ostlund2} S.~Rommer and S.~\"Ostlund: Phys. Rev. B {\bf 59} (1999) 10493.
\bibitem{Cardy} P.~Calabrese and J.~Cardy: J. Phys. A {\bf 42} (2009) 504005.
\bibitem{Trotter} H.F.~Trotter: Proc. Am. Math. Soc. {\bf 10} (1959) 545.
\bibitem{Suzuki} M.~Suzuki: Prog. Theor. Phys. {\bf 56} (1976) 1454.
\end{thebibliography}
\end{document}